\documentstyle[aps,preprint]{revtex}
\begin{document}
\draft
\author{R.Zh.~ SHAISULTANOV\thanks{%
Email:shaisultanov@inp.nsk.su}}
\address{Budker Institute of Nuclear Physics \\
630090, Novosibirsk 90, Russia}
\title{Backreaction in spinor QED and decoherence functional}
\date{}
\maketitle

\begin{abstract}
Using the Schwinger-Keldysh (closed time path or CTP) and Feynman-Vernon
influence functional formalisms we obtain an expression for the influence
functional in terms of Bogoliubov coefficients for the case of spinor
quantum electrodynamics. Then we derive a CTP effective action in
semiclassical approximation and its cumulant expansion. Using it we obtain a
equation for the description of the charged particle creation in electric
field and of backreaction of charged quantum fields and their fluctuations
on time evolution of this electric field. Also an intimate connection
between CTP effective action and decoherence functional will allow us to
analyze how macroscopic electromagnetic fields are ``measured'' through
interaction with charges and thereby rendered classical.
\end{abstract}


\newpage

\section{Introduction}

This paper is an extension of our previous work \cite{my} to the case of
spinor QED. We will study the quantum non-equilibrium effects of pair
creation in strong electric fields. Backreaction of pair creation on
electric field was recently discussed by Cooper, Mottola at all \cite{klu}.
They derived the semiclassical Maxwell equation, carry out its
renormalization and numerically solve it for some initial conditions in 1+1
dimensions.Their numerical results clearly exhibits the decay of the
electric field because of screening by the produced particles. We wish to
make a step further and derive a Langevin equation, taking into account a
noise from quantum matter fields. To this purpose we will use some mixture
of Schwinger-Keldysh (CTP) and Feynman-Vernon influence functional
formalisms. It is important to note here that phenomenological equations of
motion with noise term can also be derived using decoherence functional
formulation of quantum mechanics. This was done for some model quantum
systems in \cite{gh,brun}.

This paper is organized as follows: In Sec.\ref{f2} we give a brief review
of CTP functional formalism, mainly to introduce notations. All essential
details can be found in \cite{hu,w,we,wee}. In Sec.\ref{f3} we will obtain
an expression for the influence functional in terms of Bogoliubov
coefficients for the case of quantum electrodynamics with spin 1/2 charged
particles. Then in Sec.\ref{f4} we will obtain a CTP effective action in
semiclassical approximation and its cumulant expansion. It is the main
result of this paper. We will apply it for study of two interesting
problems. First this result will allow us to analyze the backreaction of
created charged particles on electric field and to derive a Langevin
equation, which take into account a noise from quantum matter fields. Then
in Sec.\ref{f5} we will use an intimate connection between CTP\ effective
action and decoherence functional \cite{q1,hu} to analyze how macroscopic
electromagnetic fields are ``measured'' through interaction with charges and
thereby rendered classical.

\section{ The Closed Time Path Functional Formalism in Quantum Field Theory}

\label{f2}

Usually in quantum field theory our interest is in obtaining the amplitudes
of transition from in-states to the out-states. But in many cases, mainly in
statistical physics, we are concerned with expectation values of physical
quantities at finite time. To solve such initial value problems Schwinger
has invented close time path (CTP) formalism.

Let us consider the expectation value of an arbitrary operator $A$:

\begin{equation}
\label{eq:wy1}<A>\left( t\right) =Tr\ \rho \left( t\right) \ A
\end{equation}

Here $\rho $ is the density matrix that describes the ( mixed ) state of the
system. The density matrix does not necessarily have to commute with the
Hamiltonian, in which case it describes a non-equilibrium state. Using
relation\quad $\rho \left( t\right) =U\left( t,0\right) \,\rho \left(
0\right) \,U^{-1}\left( t.0\right) $ where $U\left( t,0\right) $ is the
evolution operator, inserting the identity operator $1=U\left( t,T\right) $ $%
U\left( T,t\right) $ we obtain

\begin{equation}
\label{eq:qw}
\begin{array}{c}
<A>\left( t\right) =Tr
\text{ }\rho \left( 0\right) \text{ }U^{-1}\left( t,0\right) \text{ }A\text{
}U\left( t,0\right) = \\ Tr\text{ }\rho \left( 0\right) \text{ }U\left(
0,T\right) \text{ }U\left( T,t\right) \text{ }A\text{ }U\left( t,0\right)
\end{array}
\end{equation}
\

Equation (\ref{eq:qw}) can be pictured as describing the evolution of the
system from $0$ to $t$, inserting the operator $A$ , evolving further to
some large time $T$ \ (in practice, $T\rightarrow \infty $), and then
backwards from $T$ \ to $0$. The insertion of operator may be achieved by
introducing external sources coupled to the particular operator. This
suggests the definition of the CTP generating functional

\begin{equation}
\label{eq:ww}Z\left[ J_{+,}J_{-}\right] \equiv \exp iW\left[
J_{+},J_{-}\right] =Tr\ \rho \left( 0\right) U\left( 0,T,J_{-}\right)
U\left( T,0,J_{+}\right)
\end{equation}

In the path integral representation we have

\begin{equation}
\label{eq:wy2}Z\left[ J_{+},J_{-}\right] =\int D\phi _1D\phi _2D\phi \text{ }%
\left\langle \phi _1\left| \rho \right| \phi _2\right\rangle
\,\int\limits_{\phi _1}^\phi D\phi _{+}\int\limits_{\phi _2}^\phi D\phi
_{-}\exp i\int\limits_0^Tdt\left\{ L\left[ \phi _{+}\right] -L\left[ \phi
_{-}\right] +J_{+}\phi _{+}-J_{-}\phi _{-}\right\}
\end{equation}

The expectation values can be obtained as
\begin{equation}
\label{eq:wy3}\bar \phi _{+}=\frac{\delta W}{\delta J_{+}}%
\,\,\,\,\,\,\,,\,\,\,\bar \phi _{-}=-\frac{\delta W}{\delta J_{-}}
\end{equation}

Then the CTP effective action is

\begin{equation}
\label{eq:wy4}\Gamma _{CTP}\left[ \bar \phi _{+},\,\bar \phi _{-}\right]
=W\left[ J_{+},J_{-}\right] -J_{+}\bar \phi _{+}+J_{-}\,\,\bar \phi _{-}
\end{equation}

The equations of motion are
\begin{equation}
\label{eq:wy5}\frac{\delta \Gamma _{CTP}}{\delta \bar \phi _{+}}%
=-J_{+}\,\,\,,\,\,\,\,\frac{\delta \Gamma _{CTP}}{\delta \bar \phi _{-}\,}%
=J_{-}
\end{equation}

The physical situations correspond to solutions of the homogeneous equations
at$\bar \phi _{+}=\,\bar \phi _{-}$. Then equations are real and causal.

To apply this formalism to our situation we should substitute the $\phi $
field by the pair $\psi $ and $\sigma .$ We will be interested in
expectation values of $\,\psi $ only, so we do not couple the $\sigma $
field to an external source. Also we assume that the initial density matrix
factorizes $\rho =\rho _\psi \,\,\rho _\sigma $ . Then we have
\begin{equation}
\label{eq:zzz}
\begin{array}{c}
Z\left[ J_{+},J_{-}\right] =\int D\psi _1D\sigma _1D\psi _2D\sigma
_2\,\left\langle \psi _1\left| \rho _\psi \right| \psi _2\right\rangle
\,\,\left\langle \sigma _1\left| \rho _\sigma \right| \sigma _2\right\rangle
\, \\
\int D\psi D\sigma \,\int\limits_{\psi _1}^\psi D\psi
_{+}\,\int\limits_{\sigma _1^{}}^\sigma D\sigma _{+}\int\limits_{\psi
_2}^\psi D\psi _{-}\,\int\limits_{\sigma _2^{}}^\sigma D\sigma _{-}\,\exp
\,i\,\int\limits_0^T\,dt\,\left\{ {}\right. L_\psi \left[ \psi _{+}\right]
-L_\psi \left[ \psi _{-}\right] +J_{+}\,\psi _{+}-J_{-}\psi _{-}+ \\
L_\sigma \left[ \sigma _{+}\right] -L_\sigma \left[ \sigma _{-}\right]
+L_{int}\left[ \psi _{+}\,,\sigma _{+}\right] -L_{int}\left[ \psi
_{-}\,,\sigma _{-}\right] \left. {}\right\} = \\
\int D\psi _1D\psi _2\,\left\langle \psi _1\left| \rho _\psi \right| \psi
_2\right\rangle \,\int D\psi \,\int\limits_{\psi _1}^\psi D\psi
_{+}\,\int\limits_{\psi _2}^\psi D\psi _{-}\,\exp
\,i\,\int\limits_0^T\,dt\,\left\{ {}\right. L_\psi \left[ \psi _{+}\right]
-L_\psi \left[ \psi _{-}\right] + \\
J_{+}\,\psi _{+}-J_{-}\psi _{-}\left. {}\right\} \,\Phi \left[ \psi
_{+}\,,\psi _{-}\right]
\end{array}
\end{equation}

where $\Phi \left[ \psi _{+}\,,\psi _{-}\right] $ is the so called influence
functional
\begin{equation}
\label{eq:inf}
\begin{array}{c}
\Phi \left[ \psi _{+}\,,\psi _{-}\right] \equiv \exp \,i\,S_{IF}\left[ \psi
_{+}\,,\psi _{-}\right] =\int D\sigma _1D\sigma _2\,\,\left\langle \sigma
_1\left| \rho _\sigma \right| \sigma _2\right\rangle \int\limits_{\sigma
_1^{}}^\sigma D\sigma _{+}\,\int\limits_{\sigma _2^{}}^\sigma D\sigma _{-}\,
\\
\exp \,i\,\int\limits_0^T\,dt\left\{ L_\sigma \left[ \sigma _{+}\right]
-L_\sigma \left[ \sigma _{-}\right] +L_{int}\left[ \psi _{+}\,,\sigma
_{+}\right] -L_{int}\left[ \psi _{-}\,,\sigma _{-}\right] \right\} = \\
Tr\,\left[ U\left( T,0;\psi _{+}\right) \,\,\rho _\sigma \left( 0\right)
\,U^{-1}\left( T,0;\psi _{-}\right) \right]
\end{array}
\end{equation}

It is now easy to show, using (\ref{eq:zzz}) and (\ref{eq:inf}), that in
semiclassical approximation CTP effective action has the form
\begin{equation}
\label{eq:eff}\Gamma _{CTP}\left[ \psi _{+}\,,\psi _{-}\right] =S\left[ \psi
_{+}\right] -S\left[ \psi _{-}\right] +S_{IF}\left[ \psi _{+}\,,\psi
_{-}\right]
\end{equation}

{}From this relation one may derive the semiclassical equations of motion for
the expectation values of the $\psi $ field. It is worth noting that
\begin{equation}
\label{eq:efff}\Gamma _{CTP}\left[ \psi _{+}\,,\psi _{-}\right] =-\Gamma
_{CTP}^{*}\left[ \psi _{-}\,,\psi _{+}\right] \,\,\text{ and }\,\,\,\Gamma
_{CTP}\left[ \psi \,,\psi \right] \equiv 0.
\end{equation}

\section{Influence functional for spinor QED}

\label{f3}

In this section we wish to find the influence functional in terms of
Bogoliubov coefficients (as in \cite{q2} ). The influence functional have
now the form
\begin{equation}
\label{eq:u1}\Phi \left[ \,A^{\prime },A\right] =Tr\,\left[ U\left(
T,0;A^{\prime }\right) \,\,\rho _\psi \left( 0\right) \,U^{-1}\left(
T,0;A\right) \right]
\end{equation}

To obtain $U\left( T,0;A\right) $ we will use the Heisenberg equation of
motion
\begin{equation}
\label{eq:feq}i\stackrel{.}{\Psi }\text{{}}=\left[ \vec \alpha \left( -i%
\frac \partial {\partial \vec x}-e\vec A\right) +m\beta \right] \,\Psi
\end{equation}

We will choose$\,\,\,\vec A=\left( 0,0,A\left( t\right) \right) \;$and
\begin{equation}
\label{eq:wy6}\Psi \left( \vec x,t\right) =\sum\limits_{\vec p,\sigma }\frac
1{\sqrt{2\omega _0\left( \vec p\right) V}}\left\{ b_{\vec p\sigma }\left(
t\right) u_{\vec p\sigma }+d_{-\vec p\sigma }^{+}\left( t\right) \upsilon _{-%
\vec p\sigma }\right\} \,e^{i\,\vec p\,\vec x}
\end{equation}

where $b_{\vec p\sigma \text{ }}$and $d_{-\vec p\sigma }^{+}$ are usual
annihilation and creation operators for particles and antiparticles
respectively. In what follows we will set volume $V=1$. Here $\omega
_0\left( \vec p\right) \equiv $ $\omega \left( \vec p,0\right) $ with
\begin{equation}
\label{eq:wy7}\omega ^2\left( \vec p,t\right) =\vec p_{\bot }^2+\left(
p^3-e\,A\left( t\right) \right) ^2+m^2
\end{equation}

Bispinors $u_{\vec p\sigma }\,\,\,\,$and $\upsilon _{-\vec p\sigma }$
satisfy
\begin{equation}
\label{eq:v1}
\begin{array}{c}
\left[
\vec \alpha \left( \vec p-e\vec A\left( 0\right) \right) +m\beta \right] u_{%
\vec p\sigma }=\omega _0\left( \vec p\right) u_{\vec p\sigma } \\ \left[
\vec \alpha \left( \vec p-e\vec A\left( 0\right) \right) +m\beta \right]
\upsilon _{-\vec p\sigma }=-\omega _0\left( \vec p\right) \upsilon _{-\vec p%
\sigma }
\end{array}
\end{equation}

and we choose them in following form
\begin{equation}
\label{eq:v2}
\begin{array}{c}
u_{
\vec p\sigma }=\frac 1{\sqrt{2\left( \omega _0\left( \vec p\right) -\Pi
^3\left( 0\right) \right) }}\left( \stackrel{\wedge }{\Pi }\left( 0\right)
+m\right) \chi _\sigma \\ \upsilon _{-
\vec p\sigma }=\frac 1{\sqrt{2\left( \omega _0\left( \vec p\right) +\Pi
^3\left( 0\right) \right) }}\left( -\stackrel{\wedge }{\tilde \Pi }\left(
0\right) +m\right) \chi _\sigma \\ \text{where}\;\Pi ^\mu \left( t\right)
\equiv \left( \omega \left( \vec p,t\right) ,\vec p-e\vec A\left( t\right)
\right) =\left( \omega \left( \vec p,t\right) ,\vec \Pi \left( t\right)
\right) , \\ \tilde \Pi ^\mu \left( t\right) \equiv \left( \omega \left(
\vec p,t\right) ,-\vec \Pi \left( t\right) \right) \;\text{and}\,\,\,\,%
\stackrel{\wedge }{\Pi }=\Pi _\mu \gamma ^\mu \\ \chi _1=\left(
\begin{array}{c}
0 \\
1 \\
0 \\
-1
\end{array}
\right) ,\chi _2=\left(
\begin{array}{c}
1 \\
0 \\
1 \\
0
\end{array}
\right) \,\,\text{and}\,\,\,\,\,\,\alpha ^3\chi _1=\chi _1;\alpha ^3\chi
_2=\chi _2
\end{array}
\end{equation}

Then the solution of (\ref{eq:feq}) for creation and annihilation operators
has the following form
\begin{equation}
\label{eq:bgl}b_{\vec p\sigma }\left( t\right) =\alpha _{\vec p}\left(
t\right) \,b_{\vec p\sigma }\left( 0\right) +\beta _{\vec p}^{*}\left(
t\right) \,d_{-\vec p\sigma }^{+}\left( 0\right) \,\,;\,\,d_{-\vec p\sigma
}^{+}\left( t\right) =-\beta _{\vec p}\left( t\right) \,\,b_{\vec p\sigma
}\left( 0\right) +\alpha _{\vec p}^{*}\left( t\right) \,d_{-\vec p\sigma
}^{+}\left( 0\right)
\end{equation}

where the Bogoliubov coefficients $\alpha _{\vec p}\left( t\right) $ and $%
\beta _{\vec p}\left( t\right) $ obey to a system of ordinary first order
differential equations
\begin{equation}
\label{eq:wy8}\dot \alpha _{\vec p}\left( t\right) =i\,h_{\vec p}(t)\,\alpha
_{\vec p}\left( t\right) +i\,g_{\vec p}(t)\,\beta _{\vec p}\left( t\right)
\,\,;\,\,\,\dot \beta _{\vec p}(t)=i\,g_{\vec p}(t)\,\alpha _{\vec p}\left(
t\right) -i\,h_{\vec p}(t)\,\beta _{\vec p}\left( t\right)
\end{equation}

with
\begin{equation}
\label{eq:wy9}
\begin{array}{c}
\,h_{
\vec p}(t)=-\frac 1{\omega _0\left( \vec p\right) }\left( \vec \Pi \left(
t\right) \vec \Pi \left( 0\right) +m^2\right) =-\frac{m^2+\vec p_{\bot
}^2+\Pi ^3\left( t\right) \Pi ^3\left( 0\right) }{\omega _0\left( \vec p%
\right) }\,\,\,;\, \\ \,\,g_{\vec p}(t)\,=\frac 1{\omega _0\left( \vec p%
\right) }\sqrt{\omega _0^2\left( \vec p\right) -\left( \Pi ^3\left( 0\right)
\right) ^2}\,\,\,\,\left( \Pi ^3\left( t\right) -\Pi ^3\left( 0\right)
\right)
\end{array}
\end{equation}

Using last equations we can show that $\alpha _{\vec p}\left( t\right) $ and
$\beta _{\vec p}\left( t\right) $ are expressed via auxiliary functions $f_{%
\vec p}\left( t\right) $ and $\varphi _{\vec p}\left( t\right) \;$by
following relations
\begin{equation}
\label{eq:wy10}
\begin{array}{c}
\alpha _{
\vec p}\left( t\right) =\sqrt{\frac{\omega _0\left( \vec p\right) -\Pi
^3\left( 0\right) }{2\omega _0^2\left( \vec p\right) }}\left\{ i\,\frac d{dt}%
f_{\vec p}\left( t\right) +f_{\vec p}\left( t\right) \left[ \omega _0\left(
\vec p\right) -\Pi ^3\left( t\right) +\Pi ^3\left( 0\right) \right] \right\}
\\ \beta _{
\vec p}\left( t\right) =\sqrt{\frac{\omega _0\left( \vec p\right) +\Pi
^3\left( 0\right) }{2\omega _0^2\left( \vec p\right) }}\left\{ i\,\frac d{dt}%
f_{\vec p}\left( t\right) -f_{\vec p}\left( t\right) \left[ \omega _0\left(
\vec p\right) +\Pi ^3\left( t\right) -\Pi ^3\left( 0\right) \right] \right\}
\\ \alpha _{
\vec p}\left( t\right) =\sqrt{\frac{\omega _0\left( \vec p\right) +\Pi
^3\left( 0\right) }{2\omega _0^2\left( \vec p\right) }}\left\{ i\,\frac d{dt}%
\varphi _{\vec p}^{*}\left( t\right) +\varphi _{\vec p}^{*}\left( t\right)
\left[ \omega _0\left( \vec p\right) +\Pi ^3\left( t\right) -\Pi ^3\left(
0\right) \right] \right\} \\ \beta _{\vec p}\left( t\right) =\sqrt{\frac{%
\omega _0\left( \vec p\right) -\Pi ^3\left( 0\right) }{2\omega _0^2\left(
\vec p\right) }}\left\{ -i\frac d{dt}\,\varphi _{\vec p}^{*}\left( t\right)
+\varphi _{\vec p}^{*}\left( t\right) \left[ \omega _0\left( \vec p\right)
-\Pi ^3\left( t\right) +\Pi ^3\left( 0\right) \right] \right\}
\end{array}
\end{equation}

with the $f_{\vec p}\left( t\right) $ and $\varphi _{\vec p}\left( t\right) $
satisfying to equations
\begin{equation}
\label{eq:ss1}
\begin{array}{c}
\frac{d^2}{dt^2}f_{\vec p}\left( t\right) +\left[ \omega ^2\left( \vec p%
,t\right) -ie\stackrel{.}{A}\right] \,f_{\vec p}\left( t\right) =0;\,\,f_{%
\vec p}\left( t\right) =\frac{e^{-i\omega _0\left( \vec p\right) t}}{\sqrt{%
2\left( \omega _0\left( \vec p\right) -\Pi ^3\left( 0\right) \right) }}\,%
\text{ at }t\rightarrow 0\,\, \\ \frac{d^2}{dt^2}\varphi _{\vec p}\left(
t\right) +\left[ \omega ^2\left( \vec p,t\right) -ie\stackrel{.}{A}\right]
\,\varphi _{\vec p}\left( t\right) =0;\,\,\varphi _{\vec p}\left( t\right) =%
\frac{e^{i\omega _0\left( \vec p\right) t}}{\sqrt{2\left( \omega _0\left(
\vec p\right) +\Pi ^3\left( 0\right) \right) }}\,\text{ at }t\rightarrow
0\,\,
\end{array}
\end{equation}

Equations (\ref{eq:wy8}-\ref{eq:ss1}) enable one to consider the particle
creation in a homogeneous electric field $E(t)$ having an arbitrary time
dependence. For example the famous Schwinger results \cite{pair}, describing
pair creation in a constant electric field and the results, obtained earlier
for fields with fixed time dependance (see review \cite{pvs}) can be found
using this equations . Note that the number of created particles in $\vec p$
th mode is given by
\begin{equation}
\label{eq:ss3}n_{\vec p}=\left| \beta _{\vec p}\right| ^2
\end{equation}

Now we have enough formulae to find $U(t,0)$ using relations
\begin{equation}
\label{eq:ss2}
\begin{array}{c}
\,b_{
\vec p\sigma }\left( t\right) =U^{+}b_{\vec p\sigma }U=\alpha _{\vec p}\,b_{%
\vec p\sigma }+\beta _{\vec p}^{*}\,d_{-\vec p\sigma }^{+}\,\,; \\ \,\,d_{-%
\vec p\sigma }^{+}\left( t\right) =U^{+}d_{-\vec p\sigma }^{+}\,U=-\beta _{%
\vec p}\,\,b_{\vec p\sigma }+\alpha _{\vec p}^{*}\,d_{-\vec p\sigma }^{+}
\end{array}
\end{equation}

here we take for brevity $b_{\vec p\sigma }\equiv b_{\vec p\sigma }(0)\,;d_{-%
\vec p\sigma }^{+}\equiv d_{-\vec p\sigma }^{+}\left( 0\right) $.

We will skip the details of calculations and express our result in the
following form $U=\prod_{\vec p,\sigma }U_{\vec p\sigma }$ where $U_{\vec p%
\sigma }$ is (we drop the mode label)
\begin{equation}
\label{eq:u}U=S(r,\phi )\,R(\theta )
\end{equation}

where
\begin{equation}
\label{eq:ss4}S(r,\phi )=\exp \left[ r\left( e^{2i\phi
}\,b^{+}d^{+}+e^{-2i\phi }\,\,b\,d\right) \right] \,;\,R(\theta )=\exp
\left[ i\,\theta \,\left( d^{+}d+b^{+}b-1\right) \right]
\end{equation}

$S$ and $R$ are called two-mode squeeze and rotation operators respectively.
The parameters $r\,,\phi ,\theta $ are determined from the equations
\begin{equation}
\label{eq:ss5}\alpha =e^{i\,\theta }\,\cos r\,;\,\beta =e^{i\,\theta
\,-\,\,2i\,\phi }\sin r
\end{equation}

This expression for $U\left( t,0\right) $ may be useful in many situations.
We may ,for example, rather easily describe time evolution of density matrix
$\rho \left( t\right) $ from arbitrary initial $\rho \left( 0\right) $ ,
more interesting initial states are : vacuum state, thermal equilibrium and
coherent states. Recently the time evolution of density matrix at finite
temperature was considered in \cite{swed} using the functional Schrodinger
representation. In this paper we will deal only with vacuum initial state.
Applying (\ref{eq:u1}) and (\ref{eq:u}) we find that the influence
functional with vacuum as an initial state is given by
\begin{equation}
\label{eq:ef1}\Phi \left[ A^{\prime },A\right] =\prod_{\vec p}\left( \alpha
_{\vec p}^{^{\prime }*}\,\alpha _{\vec p}+\beta _{\vec p}^{^{\prime
}*}\,\beta _{\vec p}\right) ^2
\end{equation}

\section{Semiclassical CTP effective action and Langevin equation}

\label{f4}

We may now obtain for $\Gamma _{CTP\text{ }}$:
\begin{equation}
\label{eq:z1}
\begin{array}{c}
\Gamma _{CTP}\left[ A^{\prime },A\right] =S\left[ A^{\prime }\right]
-S\left[ A\right] +S_{IF}\left[ A^{\prime },A\right] \\
\text{where }\,\,\,\,\,\,\,\,\,\,\,\,\,\,\,S_{IF}\left[ A^{\prime },A\right]
=-i\ln \Phi \left[ A^{\prime },A\right] =-2i\,\sum_{\vec p}\ln \left[ \alpha
_{\vec p}^{^{\prime }*}\,\alpha _{\vec p}+\beta _{\vec p}^{^{\prime
}*}\,\beta _{\vec p}\right]
\end{array}
\end{equation}

It is useful to introduce new variables as
\begin{equation}
\label{eq:ss6}\Xi =\frac 12\left( A^{\prime }+A\right) ;\,\,\,\,\,\,\Delta
=A^{\prime }-A
\end{equation}

and define \cite{q2}
\begin{equation}
\label{eq:z2}C_n(t_1,...,t_n;\Xi _{t_1,0},...,\Xi _{t_n\,,0}]\equiv \frac 1{%
i^{n-1}}\frac{\delta ^n\,S_{IF}\left[ A^{\prime },A\right] }{\delta \Delta
\left( t_1\right) \ldots \delta \Delta \left( t_n\right) } \bigg|_{\Delta=0}
\end{equation}

The notation of $\,\,\,\,C_1(t_1;\Xi _{t_1,0}]$ means $C_{1\text{ }}$is a
function of $t_1$ and also a functional of $\Xi $ between endpoints $t_1$
and $0$. By virtue of (\ref{eq:efff}) the $C_n$'s are real quantities.

Now we can write $S_{IF}$ as a functional Taylor series which have sence for
all $\Delta $ and $\Xi $%
\begin{equation}
\label{eq:r1}
\begin{array}{c}
S_{IF}[A^{\prime },A]=\int\limits_0^\infty d\tau _1\Delta \left( \tau
_1\right) C_1(t_1;\Xi _{t_1,0}]+
\frac i2\int\limits_0^\infty d\tau _1\int\limits_0^\infty d\tau _2\Delta
\left( \tau _1\right) \Delta \left( \tau _2\right) C_2\left( \tau _1\,,\tau
_2;\Xi _{\tau _1,0}\,,\Xi _{\tau _2,0}\right] + \\ ...+\frac{i^{n-1}}{n!}%
\int\limits_0^\infty d\tau _1...d\tau _n\Delta \left( \tau _1\right)
...\Delta \left( \tau _n\right) \,C_n(\tau _1,...,\tau _n;\Xi _{\tau
_1,0},...,\Xi _{\tau _n\,,0}]+...
\end{array}
\end{equation}

Following Feynman's procedure of deriving Langevin equation \cite{fe} we
rewrite\footnote{%
Here we closely follow the reasoning of Hu and Matacz \cite{q2}.}

\begin{equation}
\label{eq:rr1}
\begin{array}{c}
\,e^{-
\frac 12\int\limits_0^\infty d\tau _1\int\limits_0^\infty d\tau _2\Delta
\left( \tau _1\right) \Delta \left( \tau _2\right) C_2(\tau _1\,,\tau _2;\Xi
_{\tau _1,0}\,,\Xi _{\tau _2,0}]+...+\frac{i^{\,\,n}}{n!}\int\limits_0^%
\infty d\tau _1...d\tau _n\Delta \left( \tau _1\right) ...\Delta \left( \tau
_n\right) C_n(\tau _1\,,...,\tau _n;\Xi _{\tau _1,0}\,,...,\Xi _{\tau
_n,0}]+...}= \\ =\int D\xi \,P\left[ \xi ,\Xi \right] \,\exp \left\{
i\int\limits_0^\infty d\tau \Delta \left( \tau \right) \xi \left( \tau
\right) \right\}
\end{array}
\end{equation}

The left hand side of (\ref{eq:rr1}) is interpreted now as a characteristic
functional of a stochastic process $\xi \left( \tau \right) $. The
probability density functional $P\left[ \xi ,\Xi \right] $ of $\xi \left(
\tau \right) $ can be obtained from a given influence functional by
inverting the functional fourier transform. The $C_n$'s are now cumulants of
colour noise $\xi \left( \tau \right) $.

We can use now $\Gamma _{CTP}\left[ A^{\prime },A\right] $ to obtain the
semiclassical Langevin equations of motion using (\ref{eq:wy5}) in following
form

\begin{equation}
\label{eq:r3}\frac{\delta \Gamma _{CTP}\left[ A^{\prime },A\right] }{\delta
\Delta \left( \tau \right) }\bigg|_{A^{\prime }=A}=0
\end{equation}

{}From (\ref{eq:r3}) we obtain the Langevin equation

\begin{equation}
\label{eq:st}\ddot A\left( t\right) =\,\,C_1(t;A_{t,0}]+\xi \left( t\right)
\end{equation}

In our case we have from (\ref{eq:z1}) and (\ref{eq:z2})
\begin{equation}
\label{eq:ss9}
\begin{array}{c}
\,\,C_1(t;\Xi _{t,0}]=-2e\int
\frac{d^3\vec p}{\left( 2\pi \right) ^3\omega _0\left( \vec p\right) }\left[
\left( p^3-eA\left( 0\right) \right) \left( \left| \alpha _{\vec p}\left(
t\right) \right| ^2-\left| \beta _{\vec p}\left( t\right) \right| ^2\right)
\right. \,\,\,- \\ \left. -
\sqrt{m^2+\vec p_{\bot }^2}\left( \alpha _{\vec p}^{*}\left( t\right) \beta
_{\vec p}\left( t\right) +\alpha _{\vec p}\left( t\right) \beta _{\vec p%
}^{*}\left( t\right) \,\,\right) \right] =-2e\int \frac{d^3\vec p}{\left(
2\pi \right) ^3\omega _0\left( \vec p\right) }\,\,\sqrt{m^2+\vec p_{\bot }^2}%
\left( \left| \,f_{\vec p}\left( t\right) \right| ^2-\left| \varphi _{\vec p%
}\left( t\right) \right| ^2\right) \\ C_2\left( \tau _1\,,\tau _2;\Xi _{\tau
_1,0}\,,\Xi _{\tau _2,0}\right] =\left\langle \xi \left( \tau _1\right) \xi
\left( \tau _2\right) \right\rangle = \\
=\sum_{
\vec p}\left[ \frac{2e\Pi ^3\left( 0\right) }{\omega _0\left( \vec p\right) }%
\alpha _{\vec p}\left( \tau _1\right) \beta _{\vec p}\left( \tau _1\right) +%
\frac e{\omega _0\left( \vec p\right) }\sqrt{m^2+\vec p_{\bot }^2}\left(
\alpha _{\vec p}\left( \tau _1\right) ^2-\beta _{\vec p}\left( \tau
_1\right) ^2\right) \right] \times \\ \times \left[ \frac{2e\Pi ^3\left(
0\right) }{\omega _0\left( \vec p\right) }\alpha _{\vec p}^{*}\left( \tau
_2\right) \beta _{\vec p}^{*}\left( \tau _2\right) +\frac e{\omega _0\left(
\vec p\right) }\sqrt{m^2+\vec p_{\bot }^2}\left( \alpha _{\vec p}^{*}\left(
\tau _2\right) ^2-\beta _{\vec p}^{*}\left( \tau _2\right) ^2\right) \right]
+\left( \tau _1\longleftrightarrow \tau _2\right)
\end{array}
\end{equation}

It is easy to show that without noise term Eq.(\ref{eq:st}) is equal to the
semiclassical Maxwell equation, obtained in \cite{klu}.Then renormalization
of (\ref{eq:st}) can be carried out as in \cite{klu} .

So we obtain the finite renormalized Langevin equation that describe the
process of pair production in a spatially homogeneous electric field and the
backreaction from this pairs on time evolution of the electric field. The
solution of the Langevin equation is beyond the scope of the present
paper.We plan to consider this solution in future.

\section{Decoherence in spinor QED}

\label{f5}

In this section we will show ,using the results of previous sections, how
the programme of decoherence \cite{gh,zu} can be applied in the context of
quantum electrodynamics in some detail. We then analyze an example where
macroscopic electromagnetic fields are ``measured'' through interaction with
charges and thereby rendered classical. This example was discussed recently
by Kiefer for scalar QED \cite{kief} using different point of view.

In the consistent or decoherent histories formulation of quantum mechanics
\cite{quant1,quant2,quant3}the complete description of a coupled $\psi
,\sigma $ system is given in terms of fine-grained histories $\psi \left(
t\right) ,\sigma \left( t\right) .$ Let us take as a coarse-graining
procedure of summing over the $\sigma $ field. In other words the $\sigma $
field play in our case the role of environment. Then the interference
effects between coarse-grained histories are measured by the decoherence
functional $D\left[ \psi ,\psi ^{\prime }\right] $. It was shown in \cite
{q1,hu,sinha} that the decoherence functional, which is the fundamental
object of the decoherent histories formulation, is connected with CTP\
effective action by following relation
\begin{equation}
\label{eq:keq}D\left[ \psi ,\psi ^{\prime }\right] =e^{i\,\Gamma
_{CTP}\,\,\left[ \psi \,\,,\,\psi ^{\prime }\right] }
\end{equation}

The coarse-grained history $\psi \left( t\right) $ can be described
classically if and only if the decoherence functional is approximately
diagonal , that is, $D\left[ \psi ,\psi ^{\prime }\right] \simeq 0$ whenever
$\psi \neq \psi ^{\prime }.$ Now after this very brief discussion of the
some aspects of decoherence,we will proceed to discuss an example where
macroscopic field strengths decohere through their interaction with
charges.We wish to consider, as an example, a macroscopic superposition of
two electric fields, one pointing upwards, and the other pointing downwards.
In the case of spinor QED we obtain from (\ref{eq:z1},\ref{eq:r1}) using
dimensional considerations and omitting unessential phase factor that
\begin{equation}
\label{eq:kk1}D\left[ A,A^{\prime }\right] =e^{i\,\Gamma _{CTP}\left[
A^{\prime },A\right] }\sim \,\,\exp \left\{ -\frac{Ve^2E^2}m%
\sum_{k=0}^\infty a_k\left( \frac E{E_0}\right) ^{2k}\right\}
\end{equation}

Here $a_{k\text{ }}$'s are numbers and $E_0=m^2/e\sim 10^{16}$ V/cm. It is
clear that,for example, at $E\approx 10^7$ V/cm ( when $E/E_0\sim 10^{-9}$ )
we can neglect all terms with $k\geq 1.$ Now after an easy calculation we
obtain
\begin{equation}
\label{eq:kk2}D\left[ A,A^{\prime }\right] \sim \exp \left\{ -\frac{%
3V\,e^2\,E^2}{2^7\pi m}\right\}
\end{equation}

Note that the interaction with the charge states leads to an exponential
suppression factor of the corresponding interference terms for the field; in
the infrared limit of $V\to \infty $ one finds exact decoherence.

Thus the programme of decoherence \cite{quant2} may successfully be applied
in the context of quantum field theory using the concepts and methods of
nonequilibrium statistical field theory.

{\bf Acknowledgments}

I would like to thank V.N.Baier for his interest in this work and valuable
comments on the text of the manuscript.

\end{document}